\documentclass[aps,prd,superscriptaddress,12pt,showpacs,notitlepage]{revtex4}
    \usepackage{graphicx}
    \usepackage{amsmath,amssymb,mathrsfs}
\usepackage{epsf}
\usepackage{epsfig}
\usepackage[font=small,skip=0pt,justification=raggedright]{caption}

\newcounter{fig}   \newcommand{\lbfig}[1]{\refstepcounter{fig}
\label{#1} }

\newcommand{\Tr}{{\rm Tr}}

\newcommand{\bea}{\begin{eqnarray}}
\newcommand{\eea}{\end{eqnarray}}
\newcommand{\be}{\begin{equation}}
\newcommand{\ee}{\end{equation}}

\newcommand{\re}[1]{(\ref{#1})}

\newcommand{\tr}{\mbox{tr}}

\newcommand{\eqn}{\begin{eqnarray}}
\newcommand{\eqnx}{\end{eqnarray}}

\begin{document}

\title{Crystal structures in generalized Skyrme model}

\author{I. Perapechka}
\affiliation{Department of Theoretical Physics and Astrophysics, BSU, Minsk 220004, Belarus}
\author{Ya. Shnir}
\affiliation{Department of Theoretical Physics and Astrophysics, BSU, Minsk 220004, Belarus\\
BLTP, JINR, Dubna 141980, Moscow Region, Russia\\
Department of Theoretical Physics, Tomsk State Pedagogical University, Russia}

\begin{abstract}
We investigate the properties of triply-periodic Skyrme crystals in the generalized
Skyrme model $\mathcal{L}_6 + \mathcal{L}_4 + \mathcal{L}_2+ \mathcal{L}_0$ with higher-derivative terms up to sixth order.
Three different symmetry breaking potential terms $\mathcal{L}_0$  are considered,
the generalized pion mass term, double vacuum potential and mixed potential. Various scenario of
phase transitions from the low density phase to the high density phase are examined for different choices of the parameters of the model.
In particular, we investigated limiting behavior of the Skyrme crystals in the truncated submodel without the Skyrme term $\mathcal{L}_4$
and/or without the $\mathcal{L}_2$ term.
We show that the Skyrme crystal may exist in the pure $\mathcal{L}_4$ and $\mathcal{L}_6$ models and investigated
the phase structure of these solutions.
Considering the near-BPS submodel, we found that there are indications of the phase transition from a low density quasi-liquid phase to the high density symmetric phase of the Skyrmionic matter.
\end{abstract}
\maketitle

\section{Introduction}

In 1961 Skyrme proposed a simple version of the nonlinear sigma model  \cite{skyrme},
which can be considered as an effective low-energy theory of pions.
It was suggested to consider baryons as topological solitons, with
identification of the baryon number and the topological charge $B$ of the field configuration. In this picture
the pions correspond to the linearized fluctuations of the baryon field, the potential term is necessary to give a mass to these
fluctuations.

The Skyrme model has received much attention during the last few decades.
There is a variety of soliton solutions constructed numerically \cite{Braaten:1989rg,Battye:1997qq,Houghton:1997kg}.
While the simplest configuration of degree $B=1$ is spherically symmetric,
the Skyrmions of higher topological degrees possess much
more complicated symmetries, they are symmetric with respect to the dihedral group $D_n$, the extended dihedral groups $D_{nh}$ or
$D_{nd}$ or even the icosahedral group $I_h$ \cite{Houghton:1997kg}. Certainly, there is a similarity with symmetries of crystals and
fullerenes. Indeed, one can arrange the Skyrmions in a crystalline
structure, that is periodic in all three space
dimensions \cite{Klebanov:1985qi,Goldhaber:1987pb,Castillejo:1989hq,Kugler:1988mu}.

A feature of the usual Skyrme model is that the soliton solutions do not saturate the topological bound,
the binding energy is relatively high. This observation does not agree with experimentally known low
binding energies of physical nuclei, the difference is more than an order of magnitude.
It was also observed that the energies of the configurations of
higher degrees slowly approaching the topological bound as $B$ increases, see e.g. \cite{Manton-book}.
Other important observation was that the increase of the pion mass
may strongly affect the structure of the multi-Skyrmion configurations \cite{Battye:2006tb}.

Recently a few modifications of the Skyrme model were proposed to improve its phenomenological predictions
and construct weakly bounded multisoliton configurations
\cite{Adam:2010fg,Sutcliffe:2011ig,Gillard:2015eia,Gudnason:2016mms}. Two possible directions there are related with
generalization of the original Skyrme model by inclusion into Lagrangian some additional terms which
are higher order in derivatives \cite{Marleau:1991jk,Neto:1994bu,Floratos:2001ih,Adam:2010fg}, or by
a non-standard choice of the potential term, which would decrease the attractive force between the Skyrmions
\cite{Gillard:2015eia,Gudnason:2016mms,Harland:2013rxa}. In the first case there is a possibility to truncate the model
to the so-called BPS-submodel \footnote{By analogy with the corresponding monopole solutions of the Yang-Mills-Higgs model,
which satisfy the Bogomolny-Prasad-Sommerfeld (BPS) equations,
the self-dual solutions of the integrable Skyrme submodel, are usually refer to as
"{\it BPS Skyrmions}". This of course is an abuse of terminology.}, which is invariant under volume preserving
diffeomorphisms \cite{Adam:2010fg,Adam:2010ds}. Multisoliton solutions of this
reduced model exactly saturate the topological bound, they may interact only elastically and the configuration in some sense resembles
the system of liquid drops. In the second case repulsive part of the potential separates the constituents of the multi-Skyrmion configuration
which resembles  a  loosely  bound  collection of  almost isolated spherically symmetric unit charge solitons.

In this paper we will study the Skyrmion crystals in the generalized model with sextic
term and with various choices of the potential.  In  particular,
we investigate what happens to the Skyrme crystal as the
system deforms away from the standard Skyrme model.
The parameters we vary are the coefficients at all terms of the extended model,
so both the usual Skyrme model and its self-dual truncation are the limiting cases.
As we shall see, there are indications of phase transition from a low density quasi-liquid phase to the high density symmetric phase of Skyrmionic matter in the reduced almost self-dual model. 

The paper is organized as follows: in the next two sections we discuss the construction and the symmetries of the generalized
Skyrme crystal.
The numerical results for the three different symmetry breaking potential terms which include
the generalized pion mass term, double vacuum potential and mixed potential, are presented in
Section IV. Here we also compare the results to the usual pattern of phase transitions in the
conventional Skyrme model and describe a numerical scheme we implement
to find the energy minimizers. The Skyrme crystals in  the submodels of the general model are discussed in Section V.
We give our conclusions, remarks and possible future directions
of development in the final section.

\section{Generalized Skyrme model}
The  general Skyrme model  is a Poincare invariant, nonlinear sigma model field theory.
The most general allowed form of it, restricted by the
condition that the corresponding Hamiltonian must be quadratic in time derivatives, is
\be
\label{lag}
\mathcal{L}_{0246}=\mathcal{L}_2+\mathcal{L}_4+\mathcal{L}_6+\mathcal{L}_0,
\ee
where $\mathcal{L}_0=m_\pi^2 \mathcal{V}$ is a potential term with parameter $m_\pi^2$ \footnote{Note that for some particular choices
of $\mathcal{V}$ the linearized excitations are massless, therefore $m_\pi$ is not always associated with the pion mass.}. The usual structure of the Skyrme model is given by the two terms
\be
\label{l2l4}
\mathcal{L}_2=\frac{a}{2}\;\Tr\left(L_\mu L^\mu\right),
\qquad \mathcal{L}_4=\frac{b}{4}\;\Tr\left(\left[L_\mu,L_\nu\right]\left[L^\mu,L^\nu\right]\right)\, ,
\ee
where $a$ and $b$ are nonnegative coupling constants and
\be
\label{cur}
L_\mu=U^\dagger \partial_\mu U
\ee
is the $\mathfrak{su}(2)$-valued left-invariant current, associated with
the $\mbox{SU}(2)$-valued scalar field $U=\sigma \cdot {\mathbb{I}} + i \pmb{\pi} \cdot \pmb{\tau}$.
It can be represented in terms of the quartet of
scalar fields $\mathbf{n}=\left(\sigma,\pmb{\pi}\right)$ restricted to the surface of the unit sphere $S^3$,
$\mathbf{n}^2 = \sigma^2+\pmb{\pi}\cdot\pmb{\pi}=1$.

The field of the model is required to satisfy the boundary condition $U({\bf x}) \to \mathbb{I}$
as ${\bf x} \to \infty$, thus the field
is a map $U: S^3 \mapsto S^3$ labeled by the topological invariant $B=\pi_3(S^3)$. Explicitly,
the winding number of the field configuration is given by
\be
\label{barchar}
\begin{split}
B=&\frac{1}{24\pi^2}\int d^3x \varepsilon^{ijk}~\tr \left[(U^\dagger \partial_i U)
(U^\dagger \partial_j U)
(U^\dagger \partial_k U)\right] = \frac{1}{24\pi^2}\int d^3x \varepsilon^{ijk}~\tr\left[L_i L_j L_k \right]\\
=&
-\frac{1}{12\pi^2}\int d^3x~ \varepsilon_{abcd}\varepsilon^{ijk}n^a \partial_i n^b
\partial_j n^c\partial_k n^d
\,  .
\end{split}
\ee
The corresponding topological current is
\be
\label{barcur}
B^\mu=\frac{1}{24\pi^2}\varepsilon^{\mu\nu\rho\sigma}\Tr\left(L_\nu L_\rho L_\sigma \right) \, .
\ee

Thus, the model \re{lag} includes a sextic term
\be
\label{l6}
\mathcal{L}_6=4\pi^4 c B_\mu B^\mu,
\ee
where $c$ is another nonnegative coupling constant. Note that the usual rescaling of the spatial coordinates
${\bf x} \to \sqrt{\frac{a}{b}}{\bf x}$ allows us to set two of the coupling constants to unity. However, in
order  to study qualitative properties of the solutions numerically, it
will be more more convenient to keep all four parameters of the model \re{lag}.

The choice of the symmetry breaking potential terms $\mathcal{L}_0$ is important in our discussion. Since we are interested
in study of limiting transition to the self-dual $\mathcal{L}_0+\mathcal{L}_6$ crystal,
we considered three related  potentials \cite{Adam:2012sa}, the generalized pion mass potential
\be
\label{onevac}
\mathcal{V}=\left(\Tr\left(\frac{\mathbb{I}-U}{2}\right)\right)^\alpha,
\ee
the double-vacuum potential
\be
\label{doublevac}
\mathcal{V}=\left(\Tr\left(\frac{\mathbb{I}+U}{2}\right)\Tr\left(\frac{\mathbb{I}-U}{2}\right)\right)^\alpha
\ee
and mixed potential
\be
\label{mixpot}
\mathcal{V}=\Tr\left(\frac{\mathbb{I}+U}{2}\right)\left(\Tr\left(\frac{\mathbb{I}-U}{2}\right)\right)^\alpha.
\ee
Here $\alpha$ is a positive constant which defines the type of asymptotic decay of the field. As $\alpha=1$, the potential
\re{onevac} is reduced to the usual pion mass potential, setting $\alpha=2$ corresponds to the massless potential considered in
\cite{Gillard:2015eia}. The potential \re{doublevac} corresponds to the so-called "new" potential in the planar Skyrme model \cite{Weidig:1998ii},
in the limiting case of the self-dual Skyrme submodel it yields the shell-like solutions \cite{Adam:2012sa}.

Thus, the stress-energy tensor of the general model \re{lag} is given by
\be
\begin{split}
T_{\mu\nu}&=a\;\Tr\left(L_\mu L_\nu-\frac{1}{2}\eta_{\mu\nu}L_\rho L^\rho\right)+b\;
\Tr\left(\left[L_\mu,L_\rho\right]\left[L_\nu,L^\rho\right]-
\frac{1}{4}\eta_{\mu\nu}\left[L_\rho,L_\sigma\right]\left[L^\rho,L^\sigma\right]\right)\\
&+8\pi^4 c\left(B_\mu B_\nu-\frac{1}{2}
\eta_{\mu\nu}B_\rho B^\rho\right)-\eta_{\mu\nu}m_\pi^2\mathcal{V},
\label{set}
\end{split}
\ee
where $\eta_{\mu\nu}=\mbox{diag}\left(-1,1,1,1\right)$ is the usual Minkowski metric. Then, the static
energy functional is defined by
\be
\label{energy}
\begin{split}
E=\int{T_{00}\;d^3 x}&=\int \biggl\{a\;\partial_i \mathbf{n}\cdot\partial^i \mathbf{n}+2b\;
\left(\left(\partial_i \mathbf{n}\cdot\partial^i \mathbf{n}\right)^2-
\left(\partial_i \mathbf{n}\cdot\partial_j \mathbf{n}\right)\left(\partial^i \mathbf{n}\cdot\partial^j \mathbf{n}\right)\right)\\
&+c \left(\varepsilon^{abcd}\partial_1 n_a\partial_2 n_b \partial_3 n_c n_d\right)^2+m_\pi^2\mathcal{V}\biggr\}d^3 x\, .
\end{split}
\ee

The energies of the Skyrmions satisfy the topological bound
\be
\label{SFbound}
E\geq E_1 \left| B\right|.
\ee
Setting the parameters of the model as $a=1$, $b=1$, $c=m_\pi=0$, we obtain
$E_{1}=24\pi^2$. However on the $\mathbb{R}^3$ space this bound cannot be saturated, the energy of the static unit charge Skyrmion in the usual Skyrme model
without both the sextic term and the potential is $1.23 E_1$.
On the other hand, truncation of the general model \re{lag} to the self-dual submodel with $a=b=0$ allows to attain
an equality in a similar topological bound for that system \cite{Adam:2010fg,Adam:2010ds}.

It is convenient to set the energy unit as the mass of the static Skyrmion in the usual Skyrme model without the sextic term and
the potential. Thus, it will be convenient for our purposes to normalize the energy by factor $1/24\pi^2$.

\section{Generalized Skyrme crystals}

It was noticed that the topological bound \re{SFbound} can be approximately saturated by the special arrangement of the
Skyrmions in an infinite, triple periodic in space configuration, the Skyrme crystal
\cite{Klebanov:1985qi,Goldhaber:1987pb,Castillejo:1989hq,Kugler:1988mu}. This configuration, which can be considered as a model of
dense nuclear matter, can be constructed imposing
periodic boundary conditions on the Skyrme field in three spacial dimensions and taking into account that
the symmetry generators must combine both the spacial and internal rotations of the Skyrmions. Indeed,
the field of the single Skyrmion can be approximated by the triplet of orthogonal dipoles and
the character of interaction between the Skyrmions depends on their relative orientation, there are
attractive and repulsive channels in the interaction of two solitons.

Let us consider a cubic cell of size $L$, where at the point with spacial
coordinates $(x,y,z)$ the Skyrme field is given by the quartet $(\sigma, \pi_1, \pi_2, \pi_3)$. The cell is a building block of the
cubic lattice with period $L$ in all directions.
The simplest, low-density case of symmetry of Skyrme crystal was
considered by Klebanov \cite{Klebanov:1985qi}. This configuration corresponds to the attractive channel of
interaction between the six nearest neighbors, twelve second nearest are in orientation of the repulsive channel.
This crystal has combined symmetry generated by:

(i) The spacial translation by $L$ along the $x$ axis combined with a rotation by $\pi$ around the twofold axis in isospin:
\be
\label{crystal-1}
(x,y,z) \to (x+L,y,z), \qquad (\sigma, \pi_1, \pi_2, \pi_3) \to (\sigma, -\pi_1, \pi_2, -\pi_3)\, ;
\ee

(ii) The spacial reflection in coordinate space combined with internal reflection of $\mathbf{n}$:
\be
\label{crystal-2}
(x,y,z) \to (-x,y,z), \qquad (\sigma, \pi_1, \pi_2, \pi_3) \to (\sigma, -\pi_1, \pi_2, \pi_3) \, ;
\ee

(iii) Simultaneous spacial and isospin rotations around a threefold axis:
\be
\label{crystal-3}
(x,y,z) \to (z,x,y), \qquad (\sigma, \pi_1, \pi_2, \pi_3) \to (\sigma, \pi_3, \pi_1, \pi_2).
\ee

However, such a simple cubic crystal is not a lowest energy configuration for a given value of the lattice period $L$.
More detailed analysis reveals that in the usual Skyrme model without the sextic term $\mathcal{L}_6$,
there are three different phases of the Skyrme crystal with different symmetries.

The body-centered cubic (\emph{bcc}) lattice of half-Skyrmions \cite{Goldhaber:1987pb} corresponds to
the higher density. This crystal has symmetries \re{crystal-1}, \re{crystal-2} and \re{crystal-3},
as well as additional symmetry with respect to a rotation by $\pi$ around an axis going through the points
$\left(0,\frac{L}{4},\frac{L}{2}\right)$ and $\left(\frac{L}{2},\frac{L}{4},0\right)$ in coordinate space,
and $\mbox{O}(4)$ chiral rotation of the field $\mathbf{n}$:
\be
\label{crystal-4}
(x,y,z) \to \left(\frac{L}{2}-z,\frac{L}{2}-y,\frac{L}{2}-x\right), \qquad (\sigma, \pi_1, \pi_2, \pi_3) \to (-\sigma, \pi_2, \pi_1, \pi_3).
\ee

The face-centered cubic (\emph{fcc}) lattice of Skyrmions \cite{Castillejo:1989hq,Kugler:1988mu}
corresponds to the lower density phase. Here the Skyrmions are placed on the vertices of a cube and more Skyrmions are placed
on the center of the faces. There are twelve nearest neighbor in the attractive channel in such a configuration.
Its symmetry transformations include both the transformations \re{crystal-2}, \re{crystal-3}, and  a rotation around a fourfold axis in space,
combined with the internal $SO(3)$ rotation of the pion field:
\be
\label{crystal-5}
(x,y,z) \to (x,z,-y), \qquad (\sigma, \pi_1, \pi_2, \pi_3) \to (\sigma, \pi_1, \pi_3, -\pi_2)
\ee
as well as a translation from the corner of a cube to the center of a face combined with an
$\mbox{SO}(3)$, isospin rotation acting on $\pmb{\pi}$:
\be
\label{crystal-6}
(x,y,z) \to (x+L,y+L,z), \qquad (\sigma, \pi_1, \pi_2, \pi_3) \to (\sigma, -\pi_1, -\pi_2, \pi_3).
\ee

However, the global minimum of energy of the Skyrme crystal corresponds to the
medium-density simple cubic (\emph{sc}) lattice of half-Skyrmions \cite{Kugler:1988mu}. This phase is characterized
by a spacial translation combined with an $\mbox{SO}(4)$ chiral rotation by $\pi$ in the $\sigma,\pi_1$ plane:
\be
\label{crystal-7}
(x,y,z) \to (x+L,y,z), \qquad (\sigma, \pi_1, \pi_2, \pi_3) \to (-\sigma, -\pi_1, \pi_2, \pi_3),
\ee
which replaces the transformation \re{crystal-6}; however the
transformations \re{crystal-2}, \re{crystal-3} and \re{crystal-5} are also symmetries of this phase.
The minimal energy of the usual Skyrme crystal in the $\mathcal{L}_2+\mathcal{L}_4$ model
is of $E=1.036$ at $L=4.71$ in the rescaled units of energy and length.

Note that since the symmetry group of low-density phase is a subgroup of symmetry group of medium-density phase,
the phase transition between the \emph{fcc} and medium density \emph{sc} phases should be of the second order.
Similarly, the transition between the Klebanov \emph{sc} low-density phase of single Skyrmions and
the \emph{bcc} high-density phase is of second order while the transition from the
low-density \emph{sc} or \emph{fcc} phases to the \emph{bcc} lattice is of the first order.

In order to understand the situation better, let us consider
a single Skyrmion placed at $(0,0,0)$, where $\pi_i=0$ and $\sigma =-1$. The restriction of the
reflection symmetry \re{crystal-2} together with the translational invariance \re{crystal-1} means that
$\sigma=0$ on any surface $(\pm L,\pm L,\pm L)$. A cube of side length $L$ bounded by these surfaces contains
half-Skyrmion with $\sigma < 0$. The symmetry restriction \re{crystal-7} also means that $\sigma=1$ at
the point $(L,0,0)$ where the second half-Skyrmion with $\sigma>0$ is located.
Each of the cubes has topological charge 1/2. The Skyrme crystal in the high-density phase
can be viewed as a construction builded from these
cubes of two types with Skyrmions appropriately internally rotated, thus this is
a system of half-Skyrmions arranged on a simple cubic lattice.

Our goal now is to study the pattern of phase transition in the general Skyrme crystal \re{lag}.
Note that direct minimization of the corresponding energy functional needs a large amount of computational power, thus
to simulate the crystal numerically, we follow the approach of the paper \cite{Kugler:1988mu}.
We expand unnormalized Skyrme field $\overline{\mathbf{n}}$ in a Fourier series possessing
required symmetries and then minimize the energy with respect to the coefficients of the expansion.
The normalized Skyrme field then can be recovered as
$$
\mathbf{n} = \frac{\overline{\mathbf{n}}}{\| \overline{\mathbf{n}}\| } \, .
$$

Since the general Skyrme
model \re{lag} may have different symmetries, we will look for solutions possessing only general symmetries of the crystal
\re{crystal-1}-\re{crystal-3}. For such configurations the Fourier expansion will take the form
\be
\label{fourier}
\begin{split}
\overline{\sigma}=&\sum_{a,b,c} \beta_{abc}\cos\left(\frac{a\pi x}{L}\right)\cos\left(\frac{b\pi y}{L}\right)
\cos\left(\frac{c\pi z}{L}\right),\\
\overline{\pi^1}=&\sum_{a,b,c} \alpha_{abc}\sin\left(\frac{a\pi x}{L}\right)\cos\left(\frac{b\pi y}{L}\right)
\cos\left(\frac{c\pi z}{L}\right)
\end{split}
\ee
and similarly for the components  $\overline{\pi^2}$ and $\overline{\pi^3}$ which can be obtained from $\overline{\pi^1}$ by using the
transformations \re{crystal-3} and \re{crystal-5}. Here we will take $\beta_{abc}=\beta_{bca}=\beta_{cab}$.

\section{Numerical results}
In our numerical analysis we minimize the static energy functional \re{energy} with respect to
the coefficients of the Fourier expansion \re{fourier}
using numerical optimization algorithm SNOPT \cite{SNOPT}, the relative errors are lower than $10^{-6}$.
A typical number of terms in the general Fourier expansion, which is necessary to obtain a solution is about 40,
to verify the results in some cases we extend it up to 100.
As a consistency check we also verify numerically the results reported in the
papers \cite{Klebanov:1985qi,Goldhaber:1987pb,Castillejo:1989hq,Kugler:1988mu}
for the usual Skyrme model without potential.

\begin{figure}[hbt]
\lbfig{energy-f}
\begin{center}
\includegraphics[height=.40\textheight, trim = 80 40 90 60, clip = true]{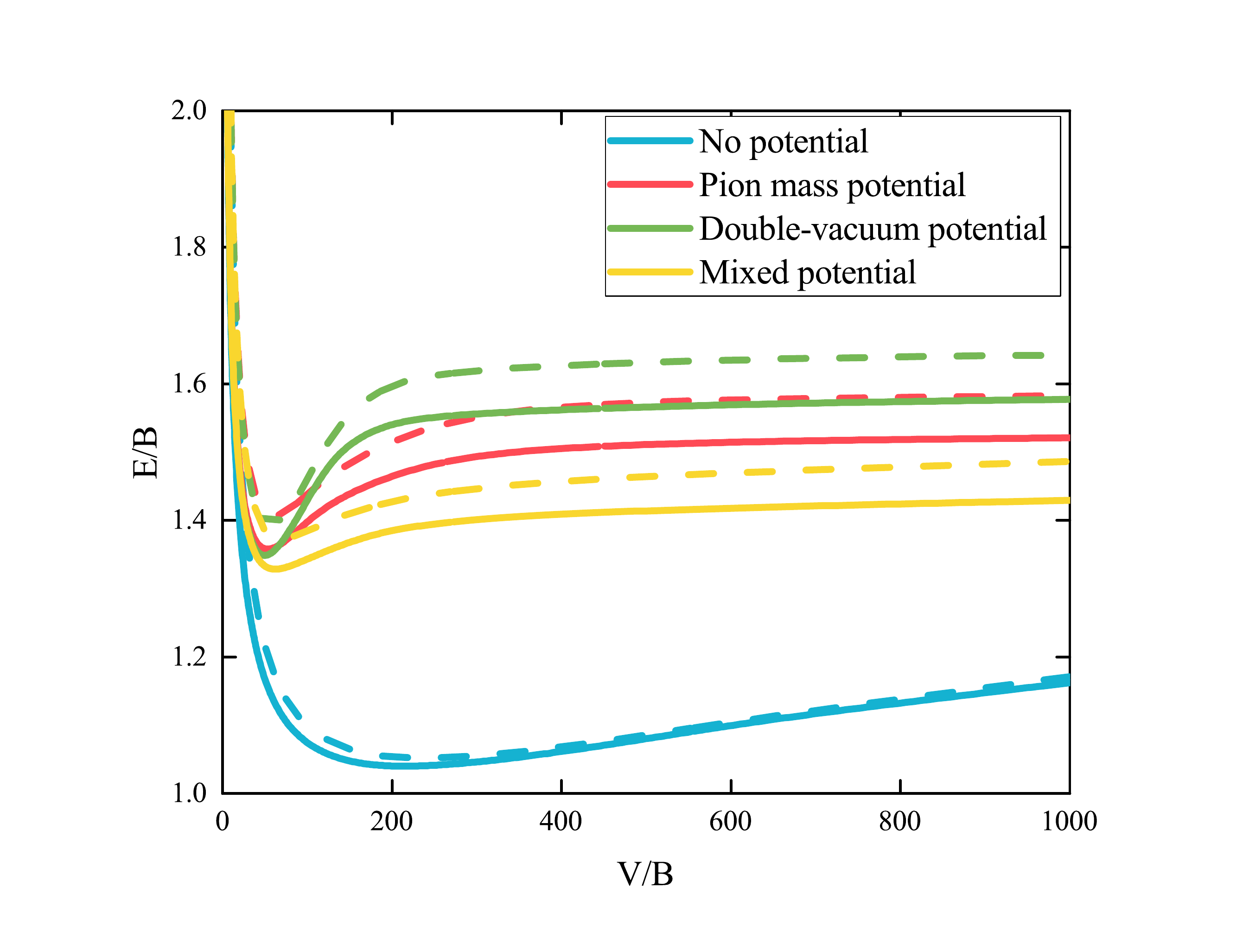}
\end{center}
\caption{\small (Color online)
Normalized energy  $E/B$ versus volume per topological charge $V/B$ for
Skyrme crystals in the general model with potentials \re{onevac},\re{doublevac} and \re{mixpot}
for $c=0$ (solid lines) and $c=1$ (dashed lines).
}
\end{figure}

First, we consider crystals in the general Skyrme model \re{lag} with different choices of the potentials
\re{onevac},\re{doublevac} and \re{mixpot}, and investigate the effect of the presence of the sextic term.
We fix the parameters of the model as $a=b=m_\pi=1$ and study dependency of the normalized energy $E/B$ of
the Skyrme crystal versus the volume of the unit cell $V/B$ for different potentials in the cases of the usual $c=0$ submodel and
$c=1$ general model. In our consideration we fix
$\alpha=1$ for the potentials \re{onevac} and \re{doublevac} and take $\alpha=2$ for the mixed potential \re{mixpot}.

\begin{figure}[hbt]
\lbfig{bcc}
\begin{center}
\includegraphics[height=.30\textheight, clip = true]{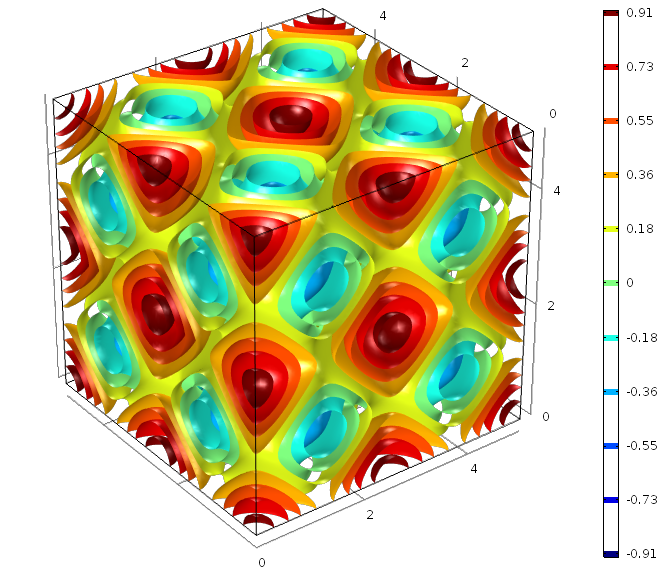}
\includegraphics[height=.30\textheight, clip = true]{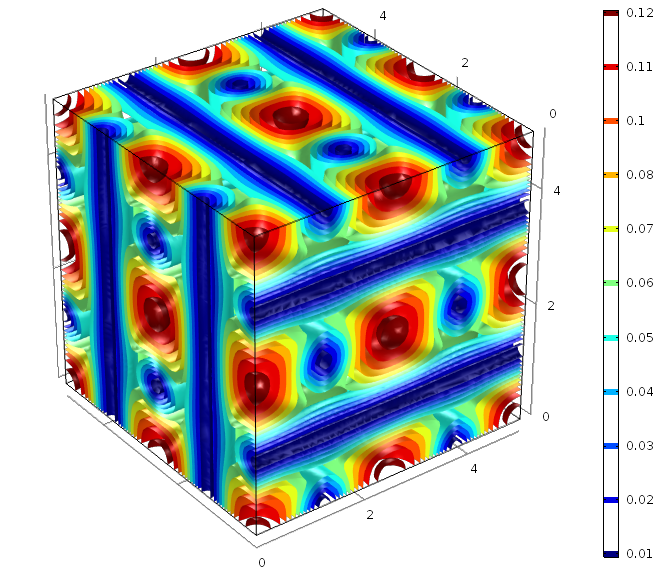}
\end{center}
\caption{\small (Color online)
Level sets of  $\sigma$ component of the Skyrme field (left panel) and
the topological charge density $B_0$ (right panel) for the \emph{bcc} Skyrmion lattice in
the model $\re{energy}$ with $a=1/2, b =1/4,
c=1$, $m_\pi=1$ and pion mass potential.
}
\end{figure}

First, we consider in detail the dependency of the normalized energy  of the crystal $E/B$ on the volume of the unit cell.
Numerical evaluation shows that in all cases these functions have a minimum at some value of the volume,
see  Fig.~\ref{energy-f}. The
energy rapidly grows as the volume of the cell tends to zero and,  as the volume  increases,
it slowly approaches the energy of
the single isolated  $B=1$ Skyrmion  in the corresponding model. As expected, the deepest minimum of the energy corresponds to the
case of the Skyrme model without the potential, the highest minimum corresponds to the model with the usual pion mass potential.
The position of the minimum also depends on the type of the potential, we found that the
the lowest density minimum corresponds to the model without the potential, as seen in the Fig.~\ref{energy-f}.
Evidently, the highest density minimum corresponds to the model with double-vacuum potential \re{doublevac}.

\begin{figure}[hbt]
\lbfig{sc}
\begin{center}
\includegraphics[height=.30\textheight, clip = true]{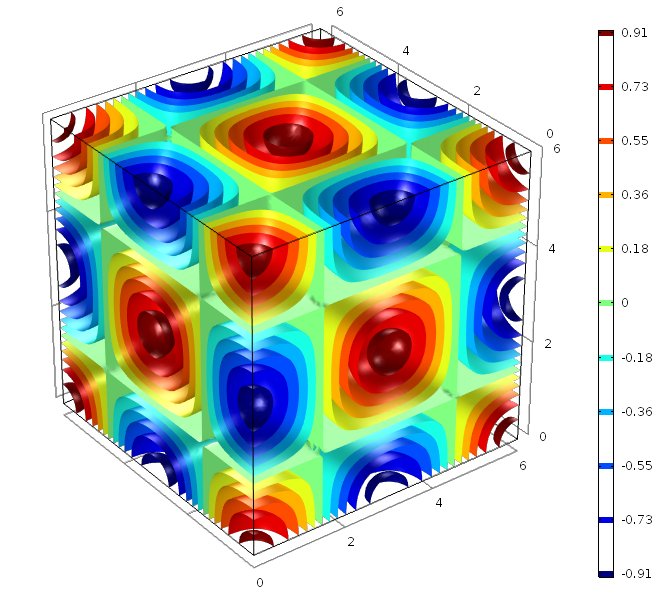}
\includegraphics[height=.30\textheight, clip = true]{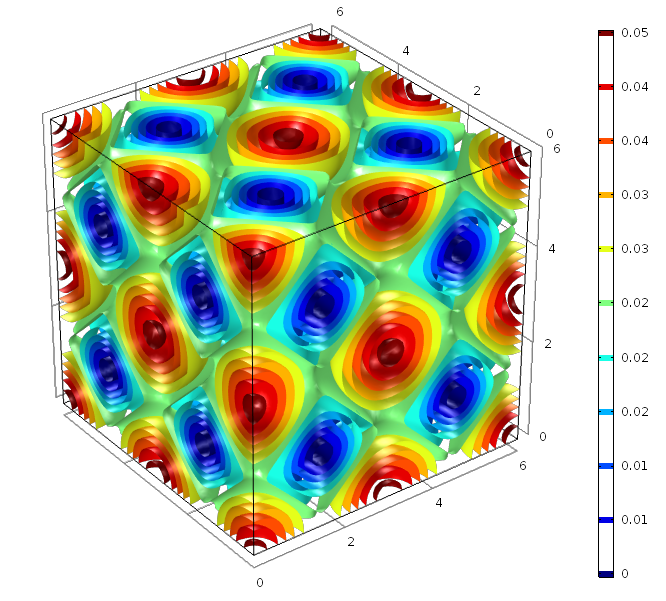}
\end{center}
\caption{\small (Color online)
Level sets of $\sigma$ component of the Skyrme field (left panel) and
the topological charge density $B_0$ (right panel) for the \emph{sc} half-Skyrmion lattice in
the model $\re{energy}$ with $a=1/2, b =1/4,
c=1$, $m_\pi=1$ and pion mass potential.
}
\end{figure}

Our results show that the inclusion of the sextic term always increases the energy of the crystal.
Indeed, this term effectively corresponds to repulsion,
it also slightly shifts the position of the minima of the energy toward lower values of density. On the other hand,
inclusion of the potential of any type yields the opposite effect shifting the position of the corresponding minima towards
higher density, the corresponding value of the minimal energy slightly increases.

\begin{figure}[hbt]
\lbfig{fcc}
\begin{center}
\includegraphics[height=.30\textheight, clip = true]{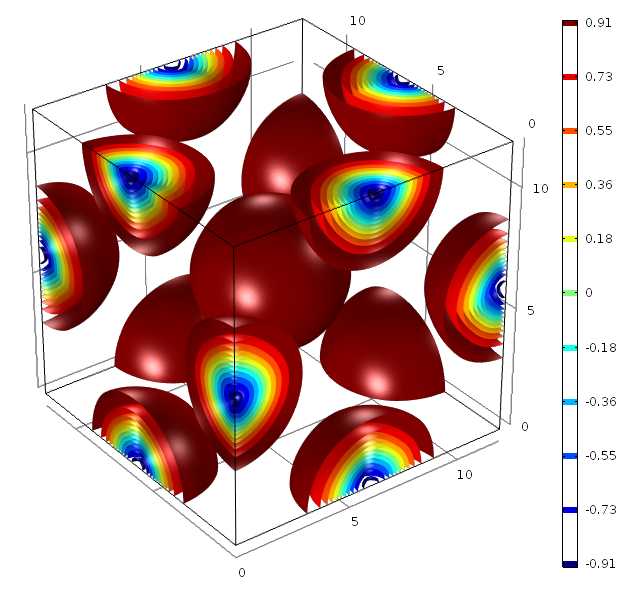}
\includegraphics[height=.30\textheight, clip = true]{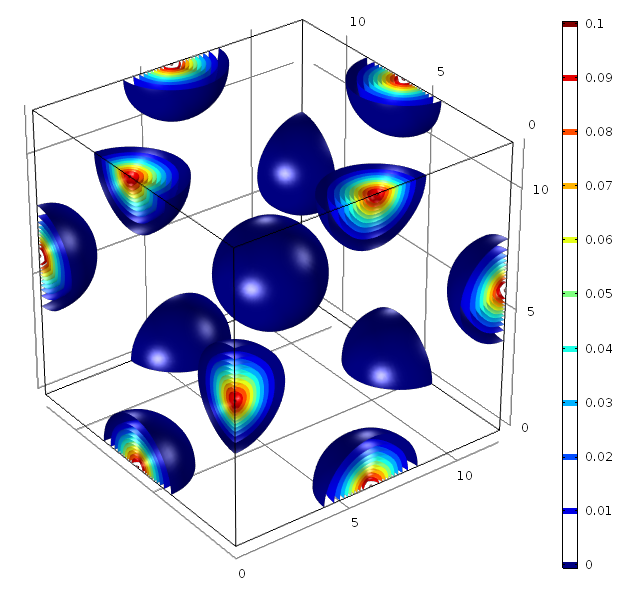}
\end{center}
\caption{\small (Color online)
Level sets of  $\sigma$ component of the Skyrme field (left panel) and
the topological charge density $B_0$ (right panel) for the \emph{fcc} Skyrmion lattice in
the model $\re{energy}$ with $a=1/2, b =1/4,
c=1$, $m_\pi=1$ and pion mass potential.
}
\end{figure}

As expected, we found that the structure of the solutions depends both on the type of the
potentials and on the density of the configurations. Recall that there are three phases of the Skyrme crystal:
the \emph{bcc}-lattice of half-Skyrmions with symmetries
\re{crystal-1}-\re{crystal-4}, which is typical for the high densities; the \emph{sc}-lattice of
half-Skyrmions with symmetries \re{crystal-2}, \re{crystal-3}, \re{crystal-5} and \re{crystal-7}, which exist at  medium
values of the density; the \emph{fcc}-lattice of Skyrmions with symmetries  \re{crystal-2}, \re{crystal-3}, \re{crystal-5},
\re{crystal-6}, which is typical for low densities.
In Figs.~\ref{bcc}, \ref{sc} and \ref{fcc} we display the typical patterns of the level sets of the $\sigma$ component of the Skyrme field
and the topological charge densities for these three types of the Skyrme crystal.

Note that in the sextic model the situation can be different, also
the structure of the minimal energy configuration depends on the presence and on the particular type of the potential term.
We have found that, in the absence of the potential of any type, the Skyrme crystal may exist in all three
phases mentioned above. Thus, as the density of the crystal increases,
two consequent phase transitions occur, one of which is of the first-order and another one is of the second-order.
The presence of the potential term affects the phases of the Skyrme crystal in different ways depending on the explicit form
of the potential.

First, we observe that the potential of any type always decreases the critical value of the volume of the unit cell,
at which the transition from the \emph{sc}-lattice to the \emph{fcc}-lattice occurs. However, it almost does not affect the critical
value of the density at which the second phase transition from \emph{fcc}-lattice, or \emph{bcc}-lattice to the high density
\emph{sc} phase of half-Skyrmion lattice occurs.

\begin{figure}[hbt]
\lbfig{phasetr}
\begin{center}
\includegraphics[height=.40\textheight, trim = 70 40 90 60, clip = true]{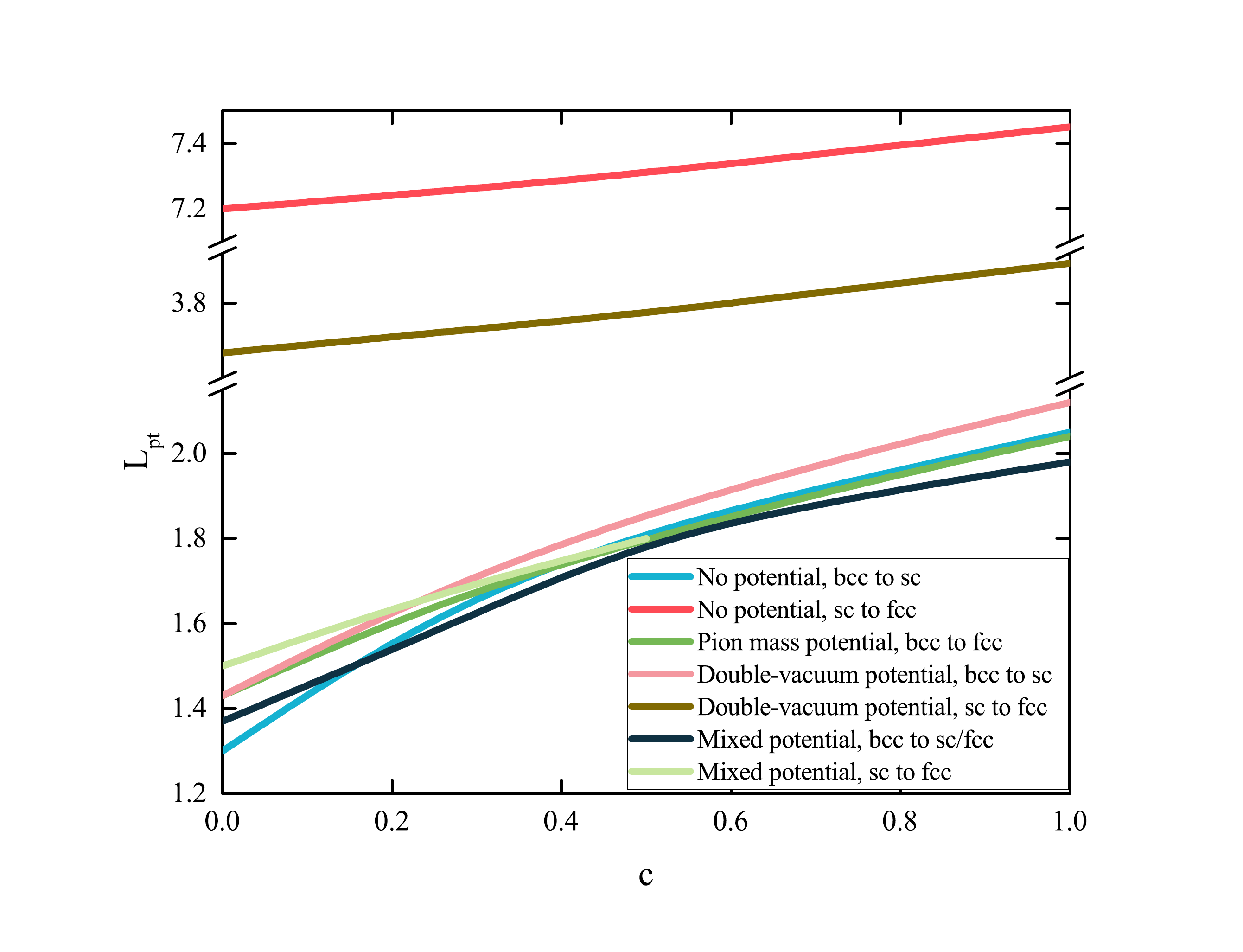}
\end{center}
\caption{\small (Color online)
Critical value of the size of the unit cell $L_{pt}$, the which the corresponding phase transition occur, as functions of
the coupling constant $c$ for all phase transitions in the general Skyrme crystal at $a=b=1$, $m_\pi=1$ and various choices of the potentials.
}
\end{figure}

Secondly, we have found that in the case of the usual pion mass potential
\re{onevac} the \emph{sc} phase of the Skyrme crystal does not exists as a global minimum for any values of the density, thus in this case
in the model with or without the sextic term, there is only one phase transition between
the \emph{bcc}-lattice and the \emph{fcc}-lattice phases.

Thirdly, we observe that in the general model with the double-vacuum potential \re{doublevac} all three phases of the
Skyrme crystal may exist. In a contrast, in the general model \re{lag} with the mixed potential \re{mixpot}, all these phases exist
only in the $c=0$ submodel without the sextic term. However, similar to the model with the usual pion mass term,
there are only two phases of the Skyrme crystal in the general case $c \neq 0$,  the \emph{bcc}- and the \emph{fcc}-lattices.

In all cases, the inclusion of the sextic term always increases the critical value of the density, at which the
corresponding phase transitions occur. In Fig.~\ref{phasetr}, we represent the results of the numerical
analysis of the dependencies of the critical value of the size of the unit cell $L_{pt}$, for which the
corresponding phase transition is taking place, on the increase of the coupling constant $c$.

\section{Skyrme crystals in the submodels of the general Skyrme model}
In this section we study
the properties of Skyrmion crystals in various submodels of the general $\mathcal{L}_{0246}$ model
with higher derivative terms \re{lag}.
Here we make use of the abbreviated notations $\mathcal{L}_{ijk}$ to label submodels of different types. Thus, in the previous
section we discussed the usual massless Skyrme model $\mathcal{L}_{24}$, the Skyrme model with potential $\mathcal{L}_{024}$ and
the submodel $\mathcal{L}_{246}$, which corresponds to the generalized theory \re{lag} without potential.

It is known that the submodel $\mathcal{L}_{06}$, where the usual Skyrme term is replaced
with a sextic term proportional to the square of the topological current, supports self-dual solutions \cite{Adam:2010fg,Adam:2010ds}.
Further, the structure of the solutions strongly depend on the explicit form of the potential \cite{Adam:2012sa,Gillard:2015eia}, in particular
for the generalized pion mass potential \re{onevac} with $\alpha < 3$ the self-dual solutions are compactons with non-analytic behavior on the
boundary of the support. Analogously, the  submodel $\mathcal{L}_{046}$ with the usual pion mass potential also supports compacton solutions.

For the sake of simplicity we further restrict our consideration to the family of models with the usual pion mass potential:
\be
\label{PMP}
\mathcal{V}=\Tr\left(\frac{\mathbb{I}-U}{2}\right).
\ee

\begin{figure}[hbt]
\lbfig{energysub}
\begin{center}
\includegraphics[height=.40\textheight, trim = 80 40 90 60, clip = true]{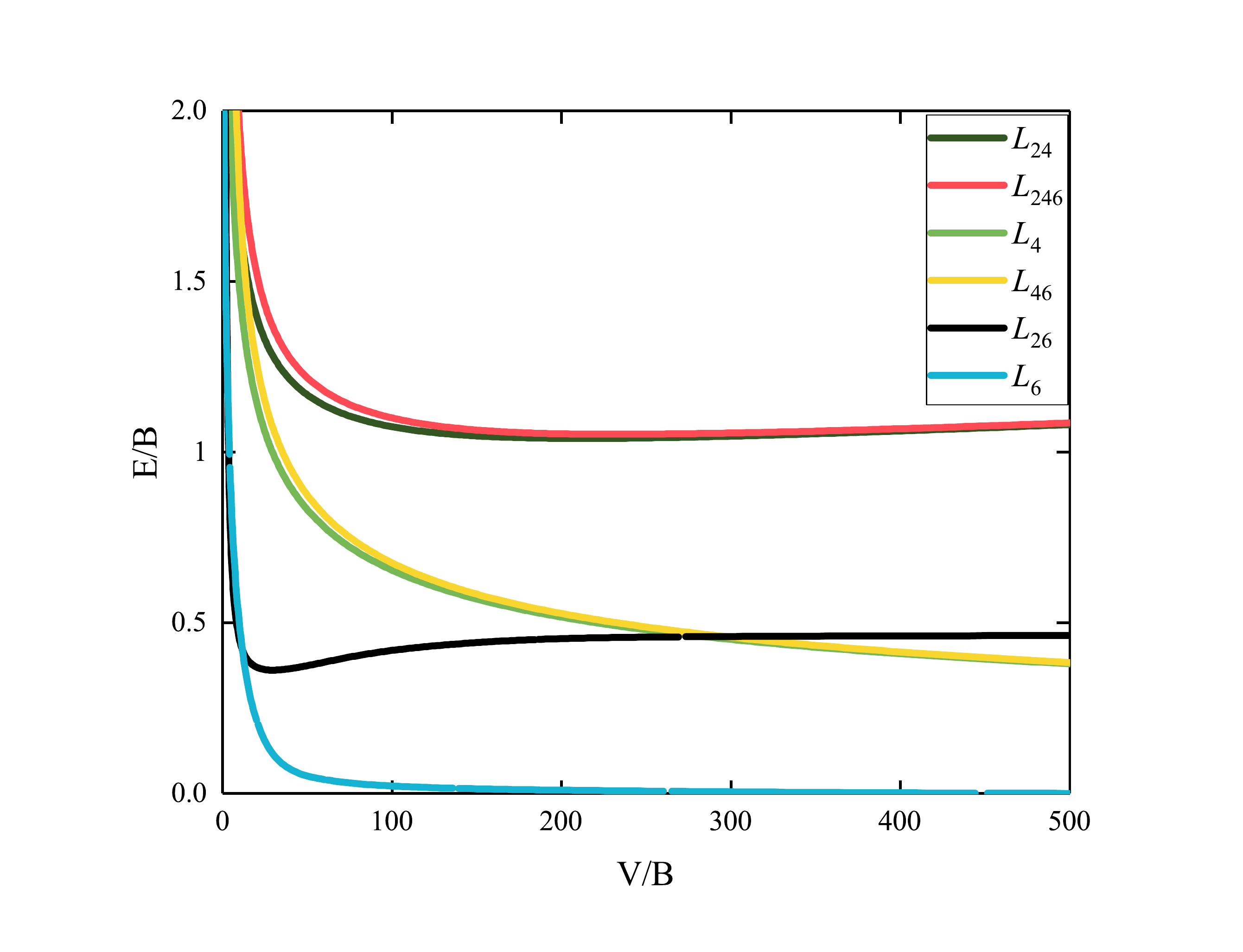}
\end{center}
\caption{\small (Color online)
Normalized energy  $E/B$ versus volume per topological charge $V/B$ for
Skyrme crystals in the general model with pion mass potential \re{PMP} and for various submodels.
}
\end{figure}

With this particular choice of the potential \re{PMP}, the solutions of the self-dual $\mathcal{L}_{06}$ submodel are
spherically-symmetric compactons \cite{Adam:2010fg,Adam:2010ds}:
\be
\label{compacton}
\sigma=\left\{
        \begin{array}{ll}
            r^2 \left(\frac{m_\pi^2}{2|B|\sqrt{c}}\right)^{\frac{1}{3}}-1 \qquad
            r\in\left[0,~~r_{cr}=\sqrt{2}\left(\frac{2|B|\sqrt{c}}{m_\pi^2}\right)^{\frac{1}{3}}\right],\\
            ~~~0~~~~~~~~~~~~~~~~~ \qquad r\geq r_{cr}
        \end{array} \right.
\ee
with the energy
\be
\label{compactonE}
E=\frac{128\pi\sqrt{2c}m_\pi^2}{15}|B|.
\ee
Thus, the solitons of the $\mathcal{L}_{06}$ submodel at zero temperature behave like a incompressible system of
noninteracting liquid droplets. Further, there is no crystal of any type in this submodel.  Indeed,
the size of the compacton $r_{cr} \sim |B|^{\frac{1}{3}}$. Therefore, the energy density distribution does not
depend on the distribution of the topological charge in the droplets. The total topological charge of the system
of non-interacting compactons is equally distributed among available volume, as the size of the cell is decreasing,
the droplets merge to decrease the energy. Finally, they form a single incompressible sphere with radius
$\sqrt{2}\left(\frac{2|B|\sqrt{c}}{m_\pi^2}\right)^{\frac{1}{3}}$ within which all the topological charge $|B|$ is concentrated.

The Skyrme crystals in the submodels $\mathcal{L}_{26}$, $\mathcal{L}_{4}$ and $\mathcal{L}_{46}$ can be considered without the potential term
since it effect is not very significant in this cases.
In  Fig.~\ref{energysub} we display dependencies of the normalized energy  $E/B$ on the volume of the unit cell for
these crystals (cf Fig.~\ref{energy-f}). Numerical evaluation shows that
for the submodel $\mathcal{L}_{26}$ without the Skyrme term, this dependency is qualitatively the same, as in the general model,
the energy per cell infinitely increases as the cell volume is decreasing to zero and it tends to the energy of
the single unit charge soliton as the volume increases. The value of the minimum of the energy is however lower, than in the full model,
it corresponds to the  higher value of density, see Fig.~\ref{energysub}.
Indeed,
it was shown by Adam and Wereszczynski \cite{Adam:2013tga} that for the  $\mathcal{L}_{26}$ submodel
the new energy bound is
\be
\quad E\geq  \frac{a}{3c} |B|
\ee
where $E$ is the normalized energy per unit cell. In our case we set $a=c=1$, thus it yields the
bound $\frac{E}{B} \geq \frac13$, while our numerical simulations gives the value of the minimal energy $E_{min} \sim 0.355 |B|$
at the lattice size $L_0=2.37$. Thus, the minimum of the energy
is just $1.5 \%$ above the corresponding energy bound for the  $\mathcal{L}_{26}$ submodel \cite{Adam:2013tga}.

Similar to the case of the usual Skyrme model, the minimal energy configuration corresponds to the same \emph{sc}-lattice of half-Skyrmions. Further, the phase structure of the Skyrme crystal in this submodel also possesses the phases we observed in the full model,
thus we may conclude that the minimal energy crystals in the $\mathcal{L}_{26}$ and $\mathcal{L}_{246}$ submodels are
quite similar for all range of values of the density.

\begin{figure}[hbt]
\lbfig{L4-L26}
\begin{center}
\includegraphics[height=.30\textheight, trim = 10 0 0 0, clip = true]{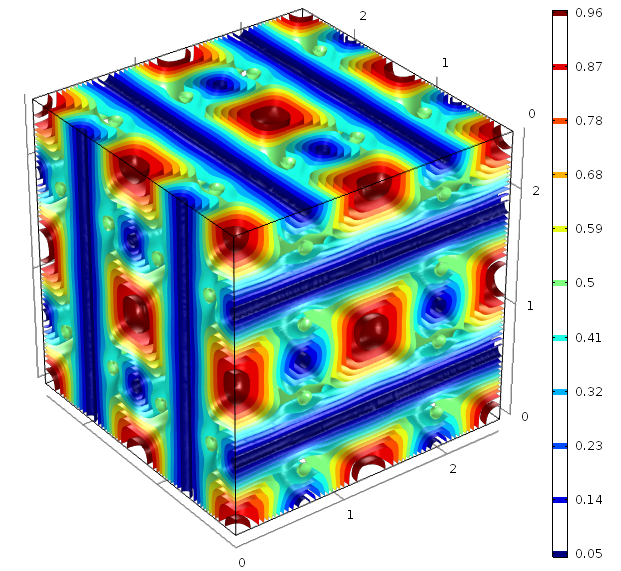}
\includegraphics[height=.30\textheight, trim = 10 0 0 0, clip = true]{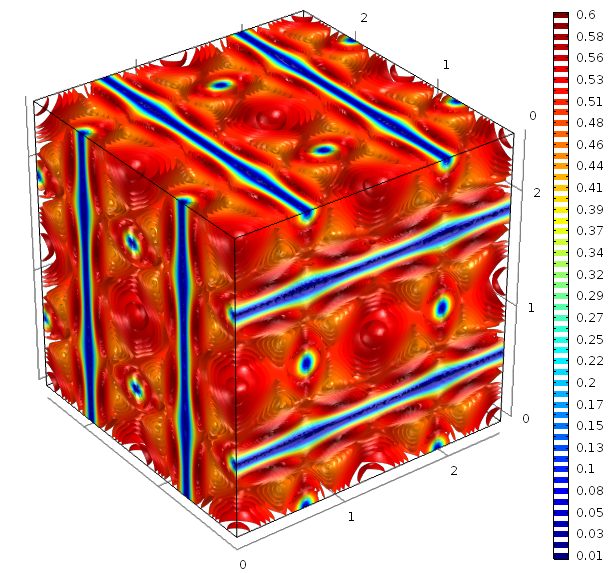}
\includegraphics[height=.30\textheight, clip = true]{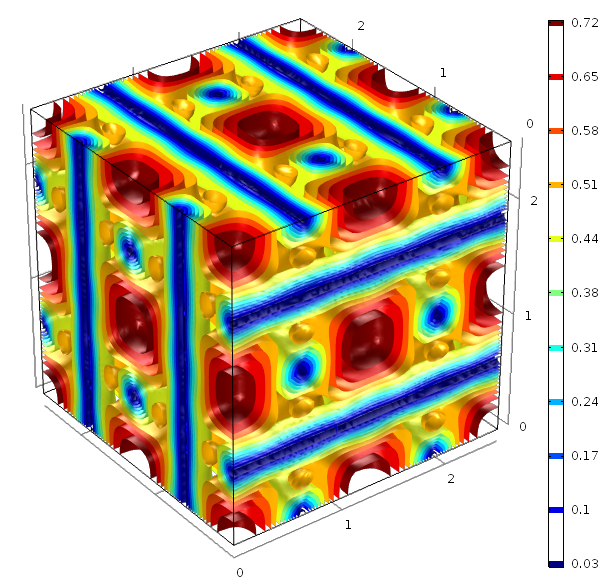}
\includegraphics[height=.30\textheight, clip = true]{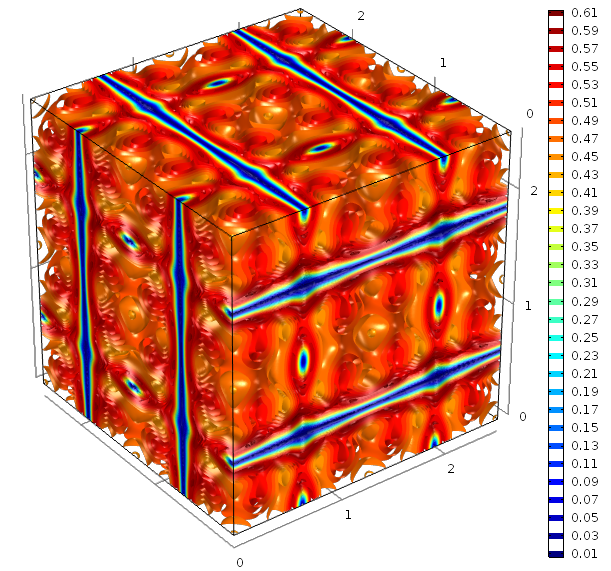}
\end{center}
\caption{\small (Color online)
Level sets of the topological charge density $B_0$ for the Skyrme crystal in the pure $\mathcal{L}_{4}$ submodel
(left upper panel), in the $\mathcal{L}_{26}$ submodel (right upper panel), in the $\mathcal{L}_{46}$ submodel (left lower panel) and in the pure $\mathcal{L}_{6}$ submodel (right lower panel) at $L=1$.
}
\end{figure}

The pattern we observed in the case of
$\mathcal{L}_{46}$ submodel is quite different from what we discussed above.
First, the dependency of the normalized energy  on the volume of the unit cell does not possess a minimum, it monotonically decreases
from infinity, in the limit of zero volume, to zero, as $L\to \infty$, see Fig.~\ref{energysub}. Indeed, there is no usual kinetic term
in such a model, a stable soliton solution cannot exist on $\mathbb{R}^3$. However, imposing the boundary condition of triple
periodicity in a crystal effectively set the theory on a torus $T^3$. It also provides a natural scale parameter of the model,
the lattice period $L$, hence the solitons may be bounded in the Skyrme crystal.
Further, it is even possible to truncate the model to the pure Skyrme $\mathcal{L}_4$ or even $\mathcal{L}_6$ crystals.
Our numerical computations show that the normalized energy of the $\mathcal{L}_4$ crystal
qualitatively depends on the volume of the unit cell in the same way, as in the $\mathcal{L}_{46}$ submodel, see Fig.~\ref{energysub}.

The second feature of the Skyrme crystals in $\mathcal{L}_{4}$ and $\mathcal{L}_{46}$ submodels above is that there we do not observe the \emph{fcc}-lattice phase.
Thus, there is only one phase transition from the  \emph{bcc}-lattice to the \emph{sc}-lattice of half-Skyrmions.
As in  the case of the full model, the presence of the sextic term yields some minor  increase of the energy of the
crystal, as well as the values of the volume of the cell, at which the normalized energy has a minimum, or the phase transition occurs. 

The peculiarity of the $\mathcal{L}_6$ crystal is that the corresponding term is strongly repulsive, the value of the
density should be rather high to stabilize the configuration. Numerical simulations indicate that this crystal may exist only in the
\emph{bcc} phase. Similar to the limiting behavior of the Skyrme crystal in the  $\mathcal{L}_{26}$ submodel, in the limit of zero volume the
normalized energy of the $\mathcal{L}_6$ crystal diverges, as it is seen in Fig.~\ref{energysub}.

For comparison we display the distribution of the topological charge density in the
$\mathcal{L}_{4}$, $\mathcal{L}_{26}$, $\mathcal{L}_{46}$ and $\mathcal{L}_{6}$ submodels at the same value of the lattice size $L=1$ in Fig.~\ref{L4-L26}. In all cases the Skyrme crystal is in the \emph{bcc}-phase, so the patterns we observe are similar. 

\subsection{Skyrme crystals in near-self-dual model}

Let us now discuss how the structure and properties of the Skyrme crystal change as the full model approaches the self-duality limit.
We consider the model \re{lag} with the pion mass potential \re{PMP} fixing the parameters $c=m_\pi=1$ and gradually decrease the other two
coupling constants. It is convenient for our purposes to set $a=b=\lambda$, $\lambda \in [0,1]$.

\begin{figure}[hbt]
\lbfig{energynearBPS}
\begin{center}
\includegraphics[height=.40\textheight, trim = 80 40 90 60, clip = true]{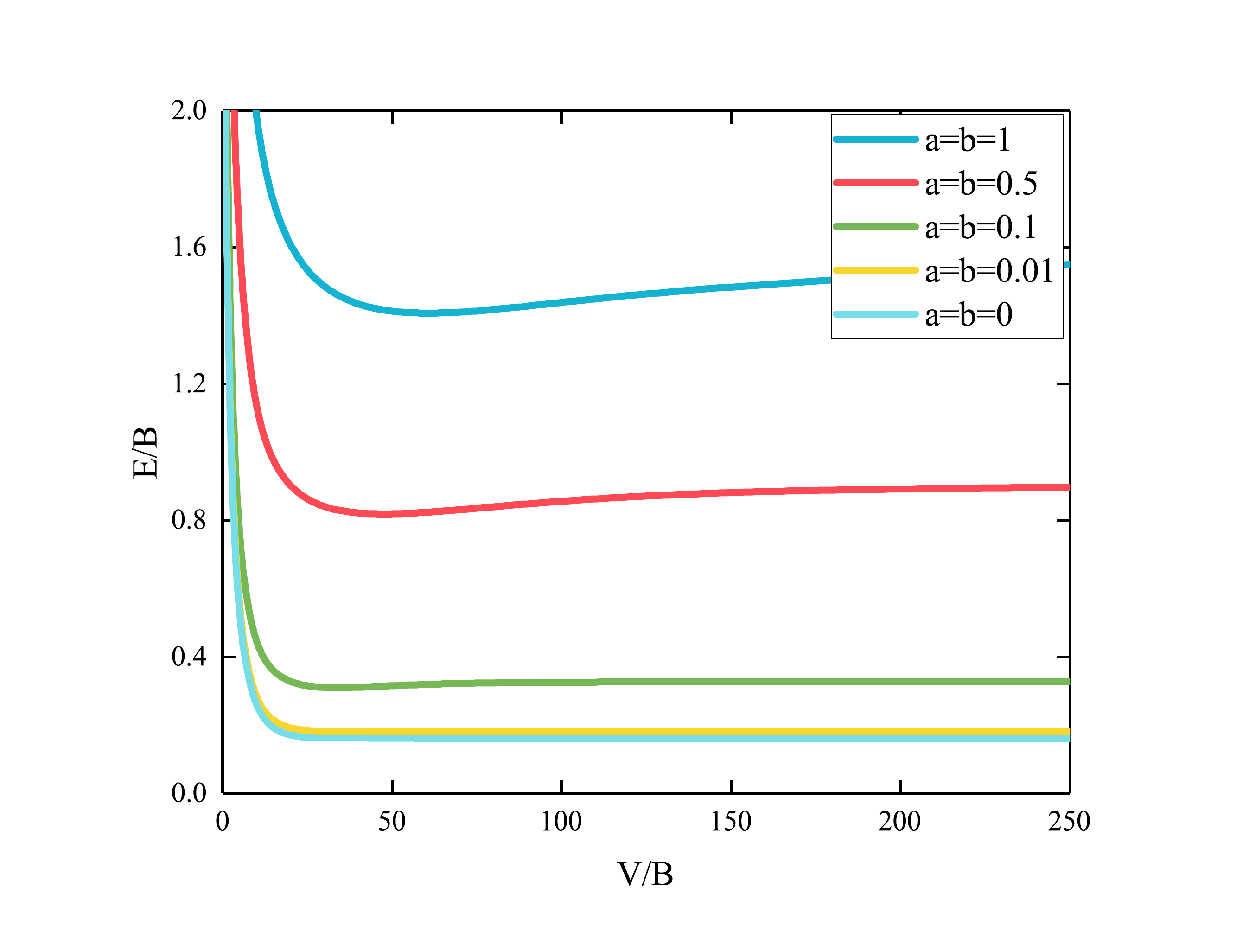}
\end{center}
\caption{\small (Color online)
Normalized energy  $E/B$ versus volume per topological charge $V/B$ for
Skyrme crystals in
near-BPS model for set of different decreasing values of $a=b$.
}
\end{figure}

Our numerical simulations shows that,  as expected, decreasing of these parameters yields significant decrease of the
value of the energy per unit cell, see Fig.~\ref{energynearBPS}. We observe that the minimal energy of the configuration
decreases almost linearly from the value $E_{min}=1.396$ at $\lambda=1$ to the minimal value $E_{min}=0.161$ at $\lambda=0$.

The second observation is that the critical value of the lattice spacing, at which the
phase transition from the \emph{bcc}-lattice of half-Skyrmions to the  \emph{fcc}-lattice of Skyrmions occurs, swiftly increases,
as the system approaches the self-duality limit. Also at the critical value of parameter $\lambda\approx0.4$ the new third phase appears.
In this phase the Skyrme crystal is transformed into the Klebanov's \emph{sc}-lattice. Recall that
the generators of symmetry of this phase are \re{crystal-1}-\re{crystal-3} and
the transition between the Klebanov's \emph{sc} low-density phase of single Skyrmions and
the \emph{bcc} high-density phase of the Skyrme crystal is of second order.
In Fig.~\ref{Klebanov}, which supplements the Figs.~\ref{bcc}-\ref{fcc}, we display the
spacial distribution of the $\sigma$ component of the Skyrme field and the topological charge density $B_0$
for such a simple cubic lattice of Skyrmions at $\lambda = 0.1$.
\begin{figure}[hbt]
\lbfig{Klebanov}
\begin{center}
\includegraphics[height=.30\textheight, clip = true]{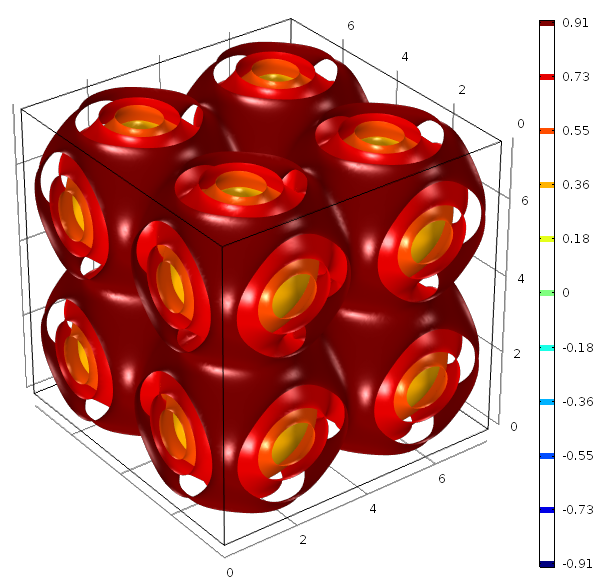}
\includegraphics[height=.30\textheight, clip = true]{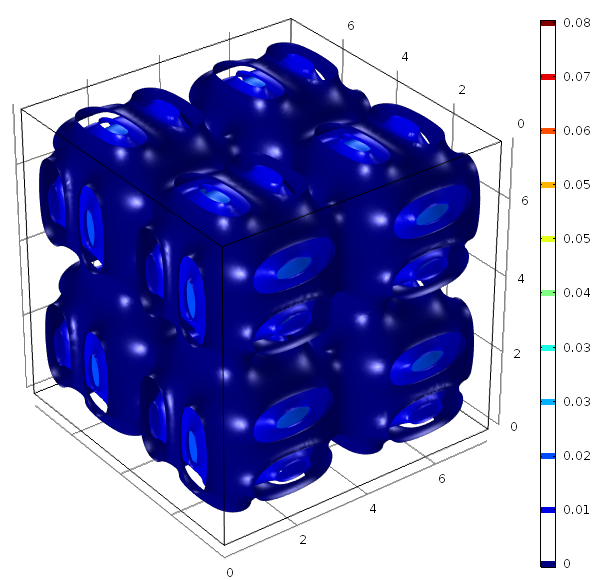}
\end{center}
\caption{\small (Color online)
Level sets of  $\sigma$ component of the Skyrme field (left panel) and
the topological charge density $B_0$ (right panel) for the \emph{sc} Skyrmion lattice in
the general $\mathcal{L}_{0246}$ model with $a=b=0.1,
c=1$, $m_\pi=1$ and the pion mass potential.
}
\end{figure}

Further, as the value of the parameter $\lambda$ continues to decrease, the contributions
to the total energy functional
\re{energy}, which come from the $\mathcal{L}_2+\mathcal{L}_4$ terms and from the $\mathcal{L}_0+\mathcal{L}_6$ terms becomes
of the same order. Then the phase transition from the \emph{sc}-lattice of Skyrmions to the  \emph{fcc}-lattice is no longer observed
and  only two phases of the Skyrme crystal remain. The \emph{fcc}-lattice phase disappears at $\lambda \approx 0.3$.
With further decrease of
$\lambda$ the  point of transition from the  \emph{bcc}-lattice to simple \emph{sc}-lattice of Skyrmions shifts towards lower values
of the density, approaching the limiting value $8r_{cr}^3$. At this point  $r_{cr}$ is just the radius of the compacton
\re{compacton}. At the same time, the normalized energy per cell approaches the value of the energy of a single compacton
\re{compactonE}.
Considering the near-BPS submodel, we found that there are indications of the phase transition from a low density quasi-liquid phase to the high density symmetric phase of the Skyrmionic matter.
Finally, the solitons become localized on a compact support and the system does not represent a crystal anymore.
Instead it can be treated as an incompressible fluid on the torus $T^3$.

\section{Conclusions}
We have investigated the properties of the Skyrme crystals in the general Skyrme model with higher derivative
sextic term being the topological current squared. This model possesses various limiting cases, among which are the usual
Skyrme model
with or without the pion mass term, and the self-dual $\mathcal{L}_{06}$ submodel.
Investigating the pattern of phase transition in these crystalline systems
we found that, depending on the density of the configuration,
the general  Skyrme crystal with minimal energy may exist in four different phases, which correspond to the low-density
\emph{sc}-lattice of Skyrmions
(Klebanov's crystal),
\emph{bcc}-lattice of half-Skyrmions, the \emph{fcc}-lattice of Skyrmions, and the high-density phase of the \emph{sc}-lattice of
half-Skyrmions. The phase transition
between the phases of \emph{fcc}-lattice of Skyrmions and the \emph{sc}-lattice of half-Skyrmions is of the second order, also
the transition between the Klebanov's phase of \emph{sc}-lattice of single Skyrmions and
the \emph{bcc}-lattice is of second order whereas the transition from the
phase of \emph{fcc}-lattice of Skyrmions to the \emph{bcc} lattice half-Skyrmions is of the first order.

We found that the addition of the repulsive sextic term to the usual Skyrme model
always increases the energy of the crystal shifting
the minima of the total energy toward lower values of density.
In the presence of the potential term, the generalized minimal energy Skyrme crystal may exist in four
phases mentioned above, the Klebanov's phase appears in the model with the pion mass potential
as the contributions of the $\mathcal{L}_2$
and the $\mathcal{L}_4$ terms becomes less significant than the energy of the strongly repulsive term
$\mathcal{L}_6$ and the potential $\mathcal{L}_0$. Further, the structure of the Skyrme crystal depends also
on the type of the potential of the model, which, depending on the values of the parameters of the model,
may eliminate some of these phase transitions.

We also considered crystalline structures in the submodels of the general Skyrme model. The most interesting finding is that
the Skyrme crystal may exist in the pure $\mathcal{L}_4$ and $\mathcal{L}_6$ models.
The pattern of the phase transitions in the first of these submodel, as well as in the similar $\mathcal{L}_{46}$ submodel,
is restricted to just one transition from the  \emph{bcc}-lattice to the high density phase of \emph{sc}-lattice of half-Skyrmions and
the normalized energy monotonically decreases as the lattice spacing increases. We also verified that the energy of the
Skyrme crystals in the $\mathcal{L}_{26}$ submodel attains an absolute minimum at some critical value of the lattice spacing. This
minimum is just $1.5 \%$ above the corresponding energy bound for this submodel \cite{Adam:2013tga}\cite{Adam:2015lra}.

Considering the near-BPS submodel, we found that there are indications of the phase transition from a low density quasi-liquid phase to the high density symmetric phase of the Skyrmionic matter.

As a direction for future work, it would be interesting
to study the generalized Skyrme crystals at finite temperature, the pattern of phase transitions in this
case can be very different. It might be also interesting to consider generalized Skyrme lattices with hexagonal symmetry,
considered by Battye and Sutcliffe in the usual Skyrme model \cite{Battye:1997wf}.

\section*{Acknowledgements}
We would like to thank  A.~Wereszczynski and C.~Adam for relevant discussions and suggestions.
I.P. would like to thank his Masha for being patient and understanding during completion of this work.
Y.S. gratefully
acknowledges support from the Russian Foundation for Basic Research
(Grant No. 16-52-12012), the Ministry of Education and Science
of Russian Federation, project No 3.1386.2017, and DFG (Grant LE 838/12-2).
He would like to thank H.~Nicolai and the staff of the AEI Golm  for
hospitality and support during the completion of this work.

\end{document}